\documentclass[10pt]{iopart}
\pdfoutput=1
\usepackage{graphicx}

\usepackage{iopams}
\usepackage{comment}
\usepackage{xcolor}

\begin{document}

%\catchline{}{}{}{}{}
%%%%%%%%%%%%%%%%%%%%%%%%%%%%%%%%%%%%%%%%%%%%%%%%%%%%%%%%%%%%%%%%%%%%

\title{Emergent Correlations in Gene Expression Dynamics as Footprints of Resource Competition}
\author{Priya Chakraborty}
\ead{chakrabortypriya95@gmail.com}
\author{Sayantari Ghosh}
\ead{sayantari.ghosh@phy.nitdgp.ac.in}
\vspace{10pt}
\address{Department of Physics, National Institute of Technology, Durgapur-713209, India}
%\cortext[cor1]{*Corresponding author}
%\begin{history}
%\received{Day Month Year}
%\revised{Day Month Year}
%\end{history}

\begin{abstract}
 Genetic circuits need a cellular environment to operate in, which naturally couples the circuit function with the overall functionality of gene regulatory network. To execute their functions all gene circuits draw resources in the form of RNA polymerases, ribosomes, and tRNAs. Recent experiments pointed out that the role of resource competition on synthetic circuit outputs could be immense. However, the effect of complexity of the circuit architecture on resource sharing dynamics is yet unexplored. In this paper, we employ mathematical modelling and \textit{in-silico} experiments to identify the sources of resource trade-off and to quantify its impact on the function of a genetic circuit, keeping our focus on regulation of immediate downstream proteins. We take the example of the fluorescent reporters, which are often used as protein read-outs. We show that estimating gene expression dynamics from readings of downstream protein data might be unreliable when the resource is limited and ribosome affinities are asymmetric. We focus on the impact of mRNA copy number and RBS strength on the nonlinear isocline that emerges with two regimes, prominently separated by a tipping point, and study how correlation and competition dominate each other depending on various circuit parameters. Focusing further on genetic toggle circuit, we have identified major effects of resource competition in this model motif, and quantified the observations. The observations are testable in wet-lab experiments, as all the parameters chosen are experimentally relevant.
\end{abstract}
\vspace{2pc}

\section{Introduction}
The interconnected structure of intra-cellular gene regulation is a massively big and complex network which we attempt to interpret in terms of relationships between several modules that are represented by specific sub-networks of protein cross-talk \cite{hecker2009gene,schlitt2007current}. Because of the unique construction of some of these modules, we often relate those to a precise function, considering it to be insulated up to a certain level from the rest of the network \cite{shen2002network,alon2007network}. However, these definite functionalities often get influenced by the other ongoing processes of the bigger network. Though the  entire  consequence of this coupling  could be extremely complex, study of the dynamical behaviours of a module under the influence of various other network components can give rise to insightful comprehension of the practical role of the module as well as the global dynamical state of the overall network.\\
Importance of mathematical modeling in understanding these effects is becoming extremely high, but the vast majority of models completely ignore the rest of the regulatory network while studying native as well as synthetic constructs. Many recent works have investigated the links between growth and gene expression dynamics considering growth rate as a prominent characteristic of the global state of the cell \cite{klumpp2009growth,ceroni2015quantifying,cameron2014tunable}. Effective models for cell growth has been used to interpret emergent nonlinear behaviours in existing pathways \cite{tan2009emergent,ghosh2011phenotypic,ghosh2012emergent,purcell2013towards}. However, most of these models consider the global effects, without the local circuit interactions that may arise from requirements of same cellular resources, including e.g. the sharing of nucleotides, polymerases, ribosomes and degradation machinery. Models that can deal with the coupled effects of regulation and competition which might be identified as retrograde effects, and can arise from in-circuit or intra-pathway resource trade-off, have not been sufficiently explored.\\
Resource sharing between different modules has drawn the attention of a growing community very recently \cite{cookson2011queueing,weisse2015mechanistic,kim2020trade}. Gene expression requires a continual participation of its essential machineries for  transcription and translation (transcription factor (TFs), RNA polymerases, ribosomes, tRNAs etc.) for synthesis of different proteins. Efficient and dependable allocation of this gene expression machinery to all the required protein synthesis pathways controls the proper execution of the cellular tasks from housekeeping to stress response. This balance leads to optimum fitness of the cell, as well as trustworthy performance of functional motifs. However, the measured supply of gene expression machinery for native functions might cause a depletion of resources available to the inserted synthetic circuit. As genetic constructs are becoming larger and more complex, their resource requirement and footprint on the host are becoming more prominent. Thus, there is a substantial probability that the circuit itself will suffer from lack of resources, and unprecedented competitions will unfold. While in one hand this can affect the homeostasis of the host, it also has the potential to initiate defects that ultimately disrupts the functionality of the circuit. For designing mathematically models for transcription, RNA polymerases, transcription factors and other cellular resources are generally considered as limitless, which does not reflect the actual scenario \cite{das2017effect,zabet2013effects}. At the level of translation, it is known that ribosomes are limited; the abundance of ribosomes is strongly correlated with to the cellular physiology and expression levels \cite{shah2013rate,scott2014emergence}.  This understanding of the interaction between protein synthesis and cellular resources plays a pivotal role in developing a quantitative method for robust reconstruction of protein-protein interaction and synthetic circuit functions from experimental data.\\
In this paper, we implement a mathematical methodology for identifying the major effects that can come into picture due to resource competition between regulatory and regulated proteins in the synthetic circuit of interest. To take into account of the resource competition, we propose a computationally favourable methodology that trusts on asymmetrical reaction probabilities and binding affinities. The article is organized as following: in Section 2, we outline the basic steps of this method by considering a two-gene system, one regulator and the other regulated, also coupled through resource competition. In Section 3, a multi-gene model of gene expression dynamics exhibiting a genetic toggle switch is considered as an illustration of the resource competition for more complex synthetic circuits. The modification of the phase diagram of the system is determined and emergent behaviours are pointed out to highlight the competition as a major limiting factor on circuit functionality. The boundary conditions for the faithful circuit operation are determined using a dynamical model. Section 4 contains discussions and concluding remarks.

\begin{figure}
\begin{center}
\includegraphics[width=13.2cm]{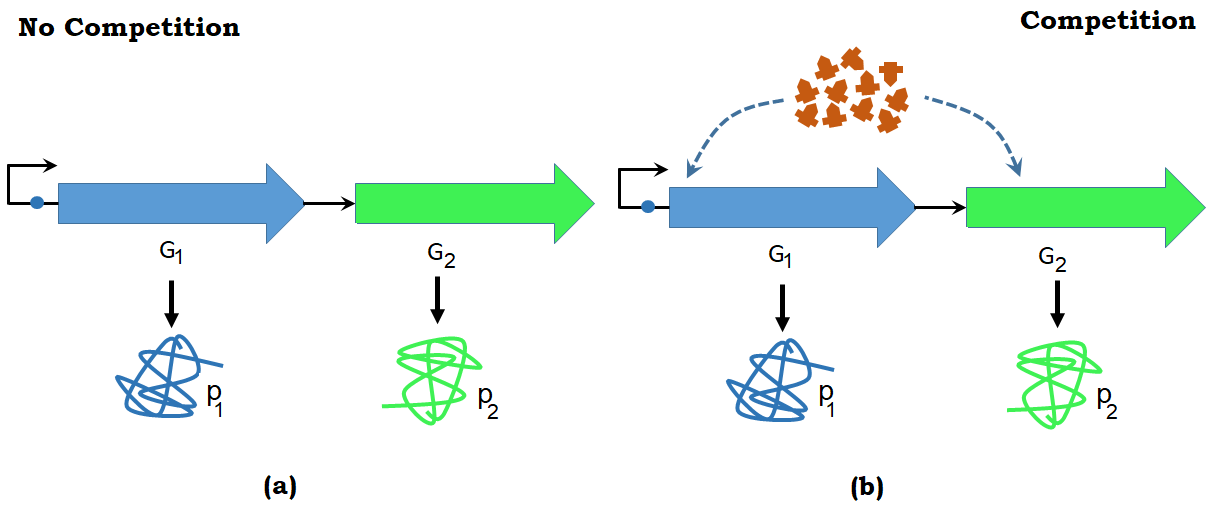}
\caption{ (a) Model 1: the regulator gene $G_1$ and associated regulated gene $G_2$ and (b) Model 2: Emergent coupling between $G_1$ and $G_2$ via resource sharing during gene expression.}
\end{center}
\label{fig:model1}
\end{figure}
\section{Dynamical Model of Resource Competition between Positively Correlated Genes}
\subsection{Model Formulation}
Let us consider a simple genetic motif to demonstrate the basic formulation of the modeling methodology. Here, we consider a gene, $G_1$ that positively regulates the expression of another gene $G_2$. Fig. \ref{fig:model1}(a) depicts the construct where the production of $p_2$ is expected to be proportional to that of $p_1$, where the $p_1$ and $p_2$ are the proteins corresponding to $G_1$ and $G_2$ respectively. Without loss of generality, this could also be considered as the case of assembling a reporter protein to a gene of interest, where the response of the reporter is supposed to be proportional to that of the gene. In the case of transcriptional fusion, the gene of interest and the reporter gene share the same promoter region; thus, the promoter activities of both the genes can be considered similar. But, due limited number of available resources at the translational level a resource competition can arise between the two genes. In our model, $G_1$ and $G_2$ which are connected by positive regulation, start competing for same resources, as shown in Fig. \ref{fig:model1}(b) We consider the resource to be ribosomes for the purpose of our model. However, these resources could be any of the components associated with the entire process of gene expression and degradation, which include ATP, transcriptional factors, ribosomes, degradation machinery etc.  Thus the model is quite flexible that can be generalized to accommodate for all these essential resources. Our assumption of considering ribosomes is based on its essential requirement for biosynthetic activities. The linear relation between ribosome concentration and growth rate \cite{zaslaver2009invariant} and similar experimental evidences, indicate that protein production is restricted by the availability of free ribosomes \cite{scott2014emergence,scott2010interdependence}. \\
\begin{figure}
\begin{minipage}[c]{0.99\linewidth}
\begin{center}
\includegraphics[width=13.2cm]{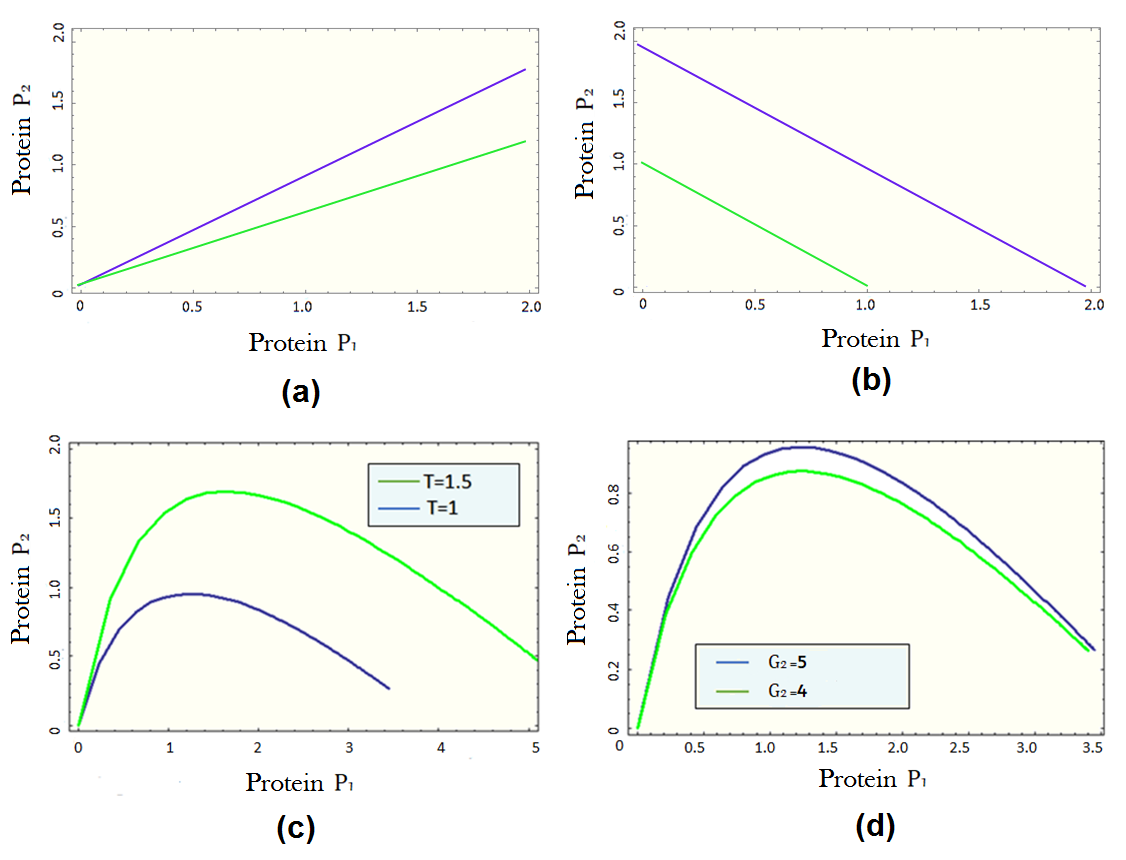}
\end{center}
\end{minipage}
\caption{Positive regulation, resource  competition and overall effect. (a) Illustration of positive regulation of protein $p_2$ by protein $p_1$. The response is expected to be linear, with the slope being dependent on strength of coupling. (b) Illustration of expression of two genes, apparently not connected, coupled through resource sharing. Limited number of resource results competition in production of protein and reflects isocost nature of falling straight lines, with slopes and intercepts dependent on available resource and copy numbers. (c) Results for the model circuit of consideration. Loss of positive correlation through an emergent Isocost-like behaviour.  $p_1-p_2$ plot keeping $G_1=G_2=5, res_2=1$ fixed and varying $res_1$ from $0$ to $10$ Green curve is when $T$ is fixed at $1.5$ and blue curve is when $T$ is fixed at $1$.  (d) Loss of positive correlation through an emergent Isocost-like behaviour.  $p_1-p_2$ plot keeping $T=1, res_2=1,G_1=5$ fixed and varying $res_1$ from $0$ to $10$.Blue curve is when $G_2$ is fixed at $5$, green is for $G_2$ fixed at $4$. }
\label{isocost}
\end{figure}
To estimate the changes in the expected output, we model the dynamical process using ordinary differential equations (ODE). We assume that $T$ is the total number of ribosomes accessible to (i.e., present in the vicinity of) the circuit of interest from the available free ribosome pool. In most bacterial cells, initiation of translation happens when the ribosome forms a complex with three initiation factors (IF-1, IF-2, and IF-3), the mRNA and initiator tRNA. The process of translation involves a number of specific non-ribosomal proteins at various stages of the translation process as well. In our model, to mathematically incorporate these steps we consider that while participating in translation of genes, the ribosomes form a complex bound state denoted by $c_1$ and $c_2$ respectively. Thus, the number of free ribosome available for translational activity, will always be $T-c_1-c_2$. Let $res_1$ and $res_2$ be the binding constant for ribosome complex formation, for the mRNAs corresponding to the two genes respectively, indicating asymmetric  affinity for ribosomes. We also consider the variability of mRNA copy numbers; for simplicity of notation in the rest of the text, for both the genes, we take $G_1$ and  $G_2$ to represent the numbers of mRNAs participating in translation from both the genes, which also takes into account of the gene copy numbers. With these considerations, the entire picture of positive regulation considering the competition effect (as shown in Fig. \ref{fig:model1}(b)) can be captured by the equations below:

\begin{equation*}
\frac{dc_1}{dt}=res_1\;(T-c_1-c_2)\;G_1-c_1\;\delta c_1
\end{equation*}
\begin{equation}
\frac{dp_1}{dt}=c_1\;\epsilon_1-p_1\;\delta p_1
\label{posreg}
\end{equation}
\begin{equation*}
\frac{dc_2}{dt}=res_2\;(T-c_1-c_2)\;G_2-c_2\;\delta c_2
\end{equation*}
\begin{equation*}
\frac{dp_2}{dt}=\frac{c_2\;\epsilon_2\;p_1}{1+p_1}-p_2\;\delta p_2
\end{equation*}
In Eq. (\ref{posreg}), $\delta c_1$,  $\delta c_2$, $\delta p_1$  and $\delta p_2$ denote the degradation rate constants for the two complexes, and the two proteins respectively. The synthesis rates of proteins from the respective complexes are given by $\epsilon_1$ and $\epsilon_2$. Here, it must be noted that we are considering that these proteins are having a positive correlation; thus the synthesis of $p_2$ has been considered to be dependent on the concentration of $p_1$  (as indicated in Fig. (\ref{fig:model1}) Thus, without the existence of any competition for resources, we implement a high (low) $p_1$ resulting into a high (low) $p_2$. The model assumes that multiple ribosomes can bind on a single mRNA. We have not considered any multimer formation for the purpose of highlighting the competition in this initial model. 
\begin{comment}
We study both these models side-by-side to understand the effect of resource competition on regulatory modules and correlated gene expression.
\end{comment}
\subsection{Loss of Positive Correlation}
\begin{figure}
\begin{minipage}[c]{0.99\linewidth}
\begin{center}
\includegraphics[width=13.2cm]{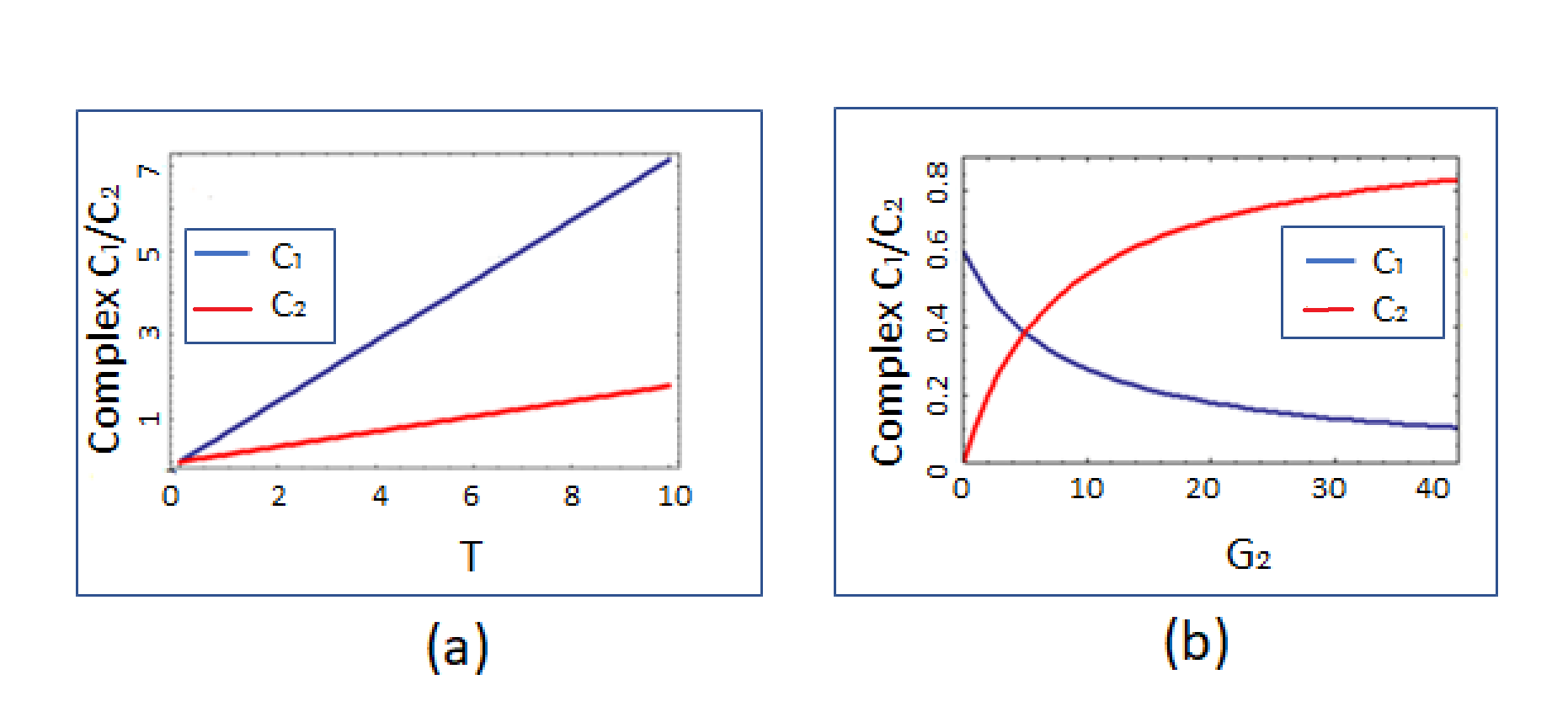}
\end{center}
\end{minipage}
\caption{Resource sharing in the light of number-based gene regulation. (a)Dependence of complex formation on free ribosome number $T$. $res_1=4,res_2=1,G_1=5,G_2=5$ is fixed and $T$ is changed from $0$ to $10$.(b) Dependence of complex formation on copy number $G_2$. The parameters $G_1=5,T=1,res_1=1,res_2=1$ is kept fixed, while $G_2$ is varied from $0$ to $40$. In both (a) and (b) the blue curve represents production of $c_1$ and the red curve is the same for $c_2$.} 
\label{stat}
\end{figure}
We analyze the model in steady state. Equating the rate of changes to zero, we note that the first and most important observation is in form of positive correlation loss for specific parameter regime.  To clarify this idea, let us define the possible limiting behaviors of this motif under study. While on the one end, this motif should reflect the positive regulation of $p_2$ by $p_1$, which we define as \textit{correlation phase}, on the other end, the limited number of resources constantly drives the system to a \textit{competition phase}. In the correlation phase the resource is sufficiently available and the circuit is under-loaded, while the  demand overload dominates the dynamics in the competition phase. Fig. \ref{isocost}(a)-(b) shows illustrative responses for both these phases. During the study of this simple construct, we expect to see the observations obeying with correlation, while the existence of competition phase is often ignored. In our model, interestingly, both these phases present their prominent appearance in the state  space, as the model shows a constant transition to competition phase from correlation phase. Fig.  \ref{isocost}(c)-(d) shows a comparative study of the behaviour of the genetic construct under different levels of competition due to asymmetrical ribosome binding affinities. For both these figures, we study the dependence of concentration of $p_2$ on that of $p_1$; for that purpose we tune the binding affinity $res_1$ keeping $res_2$ fixed at $1$.  We observe that up to a threshold affinity, the proportionality holds, but then the competition starts to dominate which draws a close similarly with isocost behavior observed in economics \cite{berndt1979engineering,gyorgy2015isocost}. The presence of positive correlation the equations gets dominated by the competition for resources, and gradually, the behaviour asymptotically converges to the  straight line resulting from  the dynamical equations that only contain the competition effects:
\begin{equation*}
\frac{dc_1}{dt}=res_1\;(T-c_1-c_2)\;G_1-c_1\;\delta c_1
\end{equation*}
\begin{equation}
\frac{dp_1}{dt}=c_1\;\epsilon_1-p_1\;\delta p_1
\end{equation}
\begin{equation*}
\frac{dc_2}{dt}=res_2\;(T-c_1-c_2)\;G_2-c_2\;\delta c_2
\end{equation*}
\begin{equation*}
\frac{dp_2}{dt}=c_2\;\epsilon_2-p_2\;\delta p_2
\label{competiton}
\end{equation*}
To investigate further, for Fig. \ref{isocost}(c), the parameter associated with gene and mRNA copy numbers, $G_1$ and $G_2$ were kept fixed at $5$, while resource affinity $res_1$ of protein $p_1$ is tuned for two different fixed values of $T$, the total available ribosomes. Increasing (decreasing) $T$ shifts the curve up (down) keeping a parallel fall due to competition which estimates the available resource `budget' has increased (decreased). In Fig. \ref{isocost}(d) we study the dependence on mRNA copy numbers, keeping $T$ and $G_1$ fixed at $1$ and $5$ respectively. We observe that in order to reach the same protein concentration of $p_2$ with less number of $G_2$ mRNA, participation of more ribosomes are needed. This consequently makes less number of ribosomes available for $p_1$ production, leading to a smaller value of $p_1$ in the competition regulated region. Thus a steeper isocost line is obtained (Fig. \ref{isocost}(d)).\\
The existence of a restricted response indicates that for asymmetrical affinity increasing concentration of $p_1$ prevents the response of $p_2$. We define $p_{1}^{th}$ as the threshold concentration of $p_1$ from which the competition phase starts to dominate; this corresponds to the turning point in Fig.  \ref{isocost}(c)-(d). Considering the maximum value of $res_1$ up to which the correlated response can be supported, it can be derived that the synthesis and degradation rate of $p_1$ controls the $p_{1}^{th}$ through the following equation:

\begin{equation}
\ \frac{p_1^2\;\delta p_1}{2}+p_1\;\delta p_1-\frac{\epsilon_1\;T}{2}=0
    \label{isoturn 1}
\end{equation}
One of the solution of Eq. (\ref{isoturn 1}) is always real and positive, where the turning occurs. Beyond  $p_1=p^{th}_{1}$, further production of $p_1$ imposes a burden on the $p_2$ synthesis; thus beyond $p^{th}_{1}$, $p_2$ starts decreasing. The maximum concentration that $p_2$ can achieve occurs at the turning point:
\begin{equation}
    p_2=\frac{\epsilon_2\;G_2\;res_2}{(\delta c_2+G_2\;res_2)\;\epsilon_1\;\delta p_2}\;\frac{p_1\;(\epsilon_1\;T-p_1\;\delta p_1)}{1+p_1}
     \label{isoturn 2}
\end{equation}
The maximum value of $res_1$ up to which the correlated response can be obtained for a fixed value of $res_2$ can also be derived in terms of the system parameters:

\begin{equation*}
    res_{1_{max}}\;=\;\frac{\delta c_2+G_2\;res_2}{G_1}\;\;\frac{\delta c_1}{\delta c_2}\;\;\sqrt{\frac{\delta p_1}{\delta p_1+T\; \epsilon_1}}
     \label{resmax}
\end{equation*}
\noindent This imposes a restrictions on the asymmetry of the resource binding affinities for observation of expected response. Interestingly, this limitation is also governed by complex and protein lifetimes.
\begin{figure}
\begin{center}
\begin{minipage}[c]{0.99\linewidth}
\includegraphics[width=13.2cm]{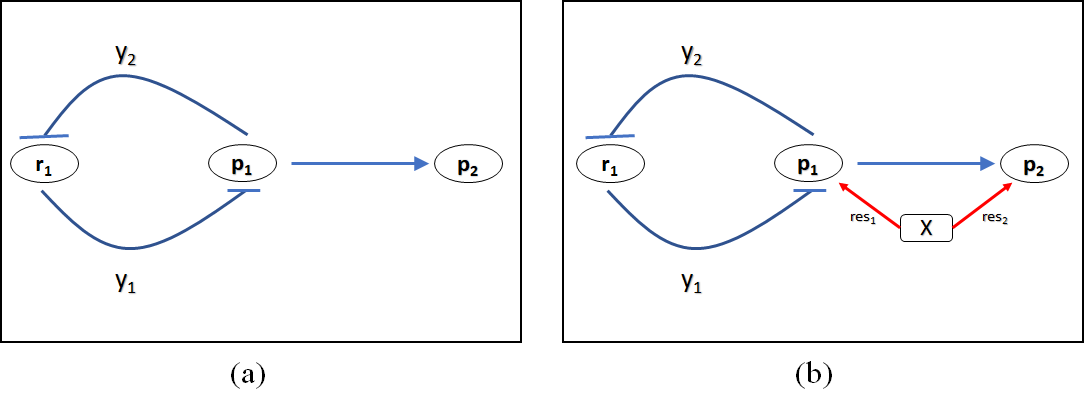}
\end{minipage}
\end{center}
\caption{(a) Genetic toggle switch with one downstream gene. The proteins $p_1$ and $r_1$ repress each other's synthesis. Repression is represented by the hammerhead symbol. (b) Genetic toggle with coupled with the downstream gene via resource competition.}
\label{togckt}
\end{figure}
\subsection{The number game of resource sharing: Statistical arguments}
In explaining the resource competition in gene expression, equilibrium statistical models of gene expression are also used as an important tool to analyze the number game of the dynamics. In this section we attempt to illustrate that our simple dynamical model can reflect several important results predicted from thermodynamic models of resource sharing and ribosome trade-off \cite{bintu2005transcriptional}. For the purpose of thermodynamic model, it is considered that the specific ribosome binding sites (with occupation number, say, $n_{on}$) are part of a pool where several other non-specific sites exist. In case of binding of ribosome to the specific site(s), the configuration is considered to be specifically bound and ready for translation of the mRNA of interest. If $n_{on}=1$ signifies a bound state, then the goal of the model is to find the probability $p(n_{on} = 1)$, which can be determined from:

\begin{equation}
    p(n_{on} = 1) =\frac{Z(n_{on}=1)}{Z_{sum}},
    \label{eq:part1}
\end{equation}
where, $Z_{sum}$ is the total partition function, i.e., the sum of all partition functions. In terms of $\varepsilon$, the energy turnover associated with the process of ribosome binding, and $D$, degeneracy of the states,  Eq. \ref{eq:part1} can be written as:
\begin{equation}
    p(n_{on} = 1) =\frac{D(n_{on}=1)\: Exp {(-\varepsilon/k_B T)}}{1+D(n_{on}=1)\: Exp {(-\varepsilon/k_B T)}},
    \label{eq:part2}
\end{equation}
This methodology has been successfully extended for translational resource sharing with the considerations of multiple ribosome binding on the same mRNA and more than one mRNA populations competing for same ribosome pool \cite{rogalla2019equilibrium}. However, the key control parameters in this modeling methodology, are the  number of ribosomes and mRNAs.\\
We test the strength of our model for predictions on a competition between two mRNA populations having statistical or thermodynamic basis \cite{rogalla2019equilibrium}. Fig. \ref{stat}(a) describes the resource competition in terms of resource affinity. Here $G_1=G_2=5$, and the affinity for resource is greater for $c_1$ than $c_2$, i.e., $(res_1=4)> (res_2=1)$. As the total number of available resources increase, $c_1$ grows much rapidly than $c_2$, which also indicates faster translation of the first mRNA population, and thus dominating production of $p_1$ protein.  When the number of competing mRNA transcripts is fixed, greater the resource affinity greater is the production of complex. \\
Next, we consider that total resource, $T$ to be fixed. Keeping the ribosome binding affinities for both the genes same ($res_1=res_2$), Fig. \ref{stat}(b) exhibits a scenario where due to resource scarcity, one mRNA gets translated at the expense of the other mRNA. Here, for $G_1=5$, we gradually vary  $G_2$ from $0$ to $40$. We note that due to absence of enough mRNA copy required for the synthesis production of $c_2$, $c_1$ is maximum and $c_2$ is $0$ initially. As $G_2$ is increased, $G_2$ must draw resource for its production from the fixed total resource pool. Allotting resource for $G_2$ will effectively decrease the availability of resource to $G_1$. Thus $c_2$ increases but $c_1$ decreases with increase of $G_2$. It is interesting to observe that there is no apparent tuning related to the parameter values associated with $c_1$  synthesis here; increasing $G_2$ increases $c_2$ but decreases $c_1$, which is the reflection of resource competition scenario. This shows that the proposed model is flexible for including mechanistic laws of gene regulation and intuitive parameters, while being consistent with the thermodynamic models that take care of number density as well as demand-supply balance.

\section{ Effects of Downstream Resource Competition on Bistable Synthetic Circuit}
\begin{figure}
\begin{center}
\includegraphics[width=11.0cm]{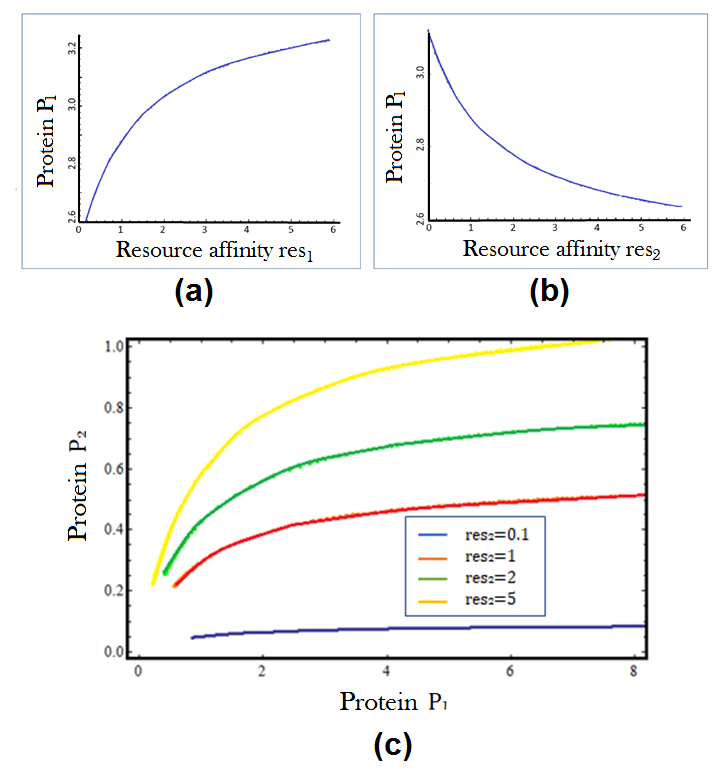}
\end{center}
\caption{Production of protein $p_1$ and $p_2$ under the limitations of resources. (a) Production of protein $p_1$ for the increase of resource affinity $res_1$,  with $res_2=1, T=1, \delta c_1= \delta c_2=2.5, \epsilon_1= \epsilon_2=1, g_1= g_2 =5, \delta p_1 = \delta p_2 =0.3, y_1=3.5, y_2 = 2$ and $res_1$ is varied from $0$ to $6$. (b) Production of protein $p_1$ for the increase of resource affinity $res_2$, with $res_1=1$ and the rest parameters are fixed as (a), while $res_2$ is varied from $0$ to $6$.  (c) Protein $p_1$ verses protein  $p_2$ plot, with $res_1=1,T=1,y_2=4,\delta c_1=\delta c_2=3.5,\epsilon_1=\epsilon_2=2,g_1=g_2=5,\delta p_1=\delta p_2=0.3$ and $y_1$ is varied from $0$ to $10$}
\label{satur}
\end{figure}
\subsection{Multigene motif: Ribosome competition for Genetic toggle}
In this section, we formulate a model for resource trade-off effects in Genetic Toggle switch, one of the most familiar motifs \cite{gardner2000construction}, occurring in natural as well as synthetic architecture commonly. Here we consider that one of the genes involved in the toggle switch, positively regulates a third downstream gene. Let us consider two protein here, $p_1$ and $r_1$, each one repressing the other with a strength $y_2$ and $y_1$ respectively, forming a genetic toggle switch. A protein $p_2$ is activated by the protein $p_1$; this $p_2$ may further activate a downstream pathway or could also be a reporter protein. The schematics of the circuit can be observed in Fig. (\ref{togckt}). From the structure  of the toggle switch it can be noted that, due to the presence of double negative feedback loop it gives rise to a bistable response in specific parameter regimes. If $p_1$ is at a  little bit higher (lower) concentration, then the feedback will drive $r_1$ to show low (high) response. Being activated by $p_1$, $p_2$ is expected to show a value linearly proportional to $p_1$. Now, let us consider a case where $p_1$ and $p_2$ are being expressed in close proximity, and thus share the resource $(X)$. As the the resource is not infinitely available, there must be a competition for resource between these two.  \\
Here we must note that $p_2$ is \textit{not} one of the regulatory proteins of the toggle circuit, rather it is a downstream protein. Because of this the effect of this competition seems apparently unimportant for the toggle functionality. We explore  these effects by the methodology discussed in  Section 2. The coupled translation  of these two mRNAs with a possibility of multiple ribosome binding are depicted through the reactions mentioned below, which, other than the mRNAs, $g_1$ and $g_2$, also involve the resource $(X)$, the bound complex of $X$ and $p_1$ $(c_1)$ and the bound complex of $X$ and $p_2$ $(c_2)$: 

\begin{tabular}{c c c c c c}
&\;$_{res_1}$& & & \;$_{res_2}$&\\
$g_1+X$ & $\rightleftharpoons$ & $g_1+c_1$ & \;\;\;\;\;\;\;\;\;\;\;\;\;\;\;\;$g_2+X$ & $\rightleftharpoons$ & $g_2+c_2$
\\ & \;$^{\delta c_1}$ & & & \;$^{\delta c_2}$&
\end{tabular}

\begin{tabular}{c c c c c c}
&$\;_{\epsilon_1}$&&&$\;_{\epsilon_2}$&\\
$c_1$ & $\longrightarrow$ & $c_1+p_1$&$\;\;\;\;\;\;\;\;\;\;\;\;\;\;\;\;\;\;\;\;\;\;\;\;c_2$ & $\longrightarrow$ & $c_2+p_2$
\end{tabular}

\begin{tabular}{c c c c c c}
&$\;_{\delta p_1}$&&&$\;_{\delta p_2}$&\\
$p_1$& $\longrightarrow$ &  $\Phi$ & $\;\;\;\;\;\;\;\;\;\;\;\;\;\;\;\;\;\;\;\;\;\;\;\;\;\;\;\;\;\;\;\;\;\;p_2$& $\longrightarrow$  & $\Phi$
\end{tabular}
\\
\\
\noindent Now the dynamical equation for total model can be written as,
\begin{equation*}
\frac{dp_1}{dt}=\frac{y_1}{1+r_1^n}-p_1+c_1\:\epsilon_1-p_1\:\delta p_1
\end{equation*}
\begin{equation*}
\frac{dc_1}{dt}=res_1\:(T-c_1-c_2)\:g_1-c_1\:\delta c_1
\end{equation*}
\begin{equation*}
\frac{dp_2}{dt}=\frac{c_2\:\epsilon_2\:p_1}{1+p_1}+\epsilon_0-p_2\:\delta p_2-p_2
\end{equation*}
\begin{equation*}
\frac{dc_2}{dt}=res_2\:(T-c_1-c_2)\:g_2-c_2\:\delta c_2
\end{equation*}
\begin{equation*}
\frac{dr_1}{dt}=\frac{y_2}{1+p_1^n}-\delta r_1\:r_1
\end{equation*}
\noindent Here, considering similar notations as before, we take $T$ as the total  amount of the resource $X$, $g_1$ and $g_2$ are the mRNA copy available for the respective complex formation, $\delta c_1$ and $\delta c_2$ are the complex degradation rates respectively of $c_1$ and $c_2$, $\delta p_1$, $\delta p_2$ and $\delta r_1$ are the protein degradation rates of $p_1$ , $p_2$ and $r_1$ respectively, $n$ is the cooperatevity (taken as $2$ throughout the analysis), $\epsilon_1$ and $\epsilon_2$ are the production rates of protein $p_1$ and $p_2$ from their corresponding complex $c_1$ and $c_2$ respectively. For generality, we have considered a basal synthesis rate, $\epsilon_0$ of $p_2$, which is small, constant and independent of protein $p_1$. As soon as the system of reactions reach steady state, time evolution of all above equations is become zero and we observe the effect of resource competition in this condition. 
\begin{figure}
\begin{minipage}[c]{0.99\linewidth}
\begin{center}
\includegraphics[width=12.2cm]{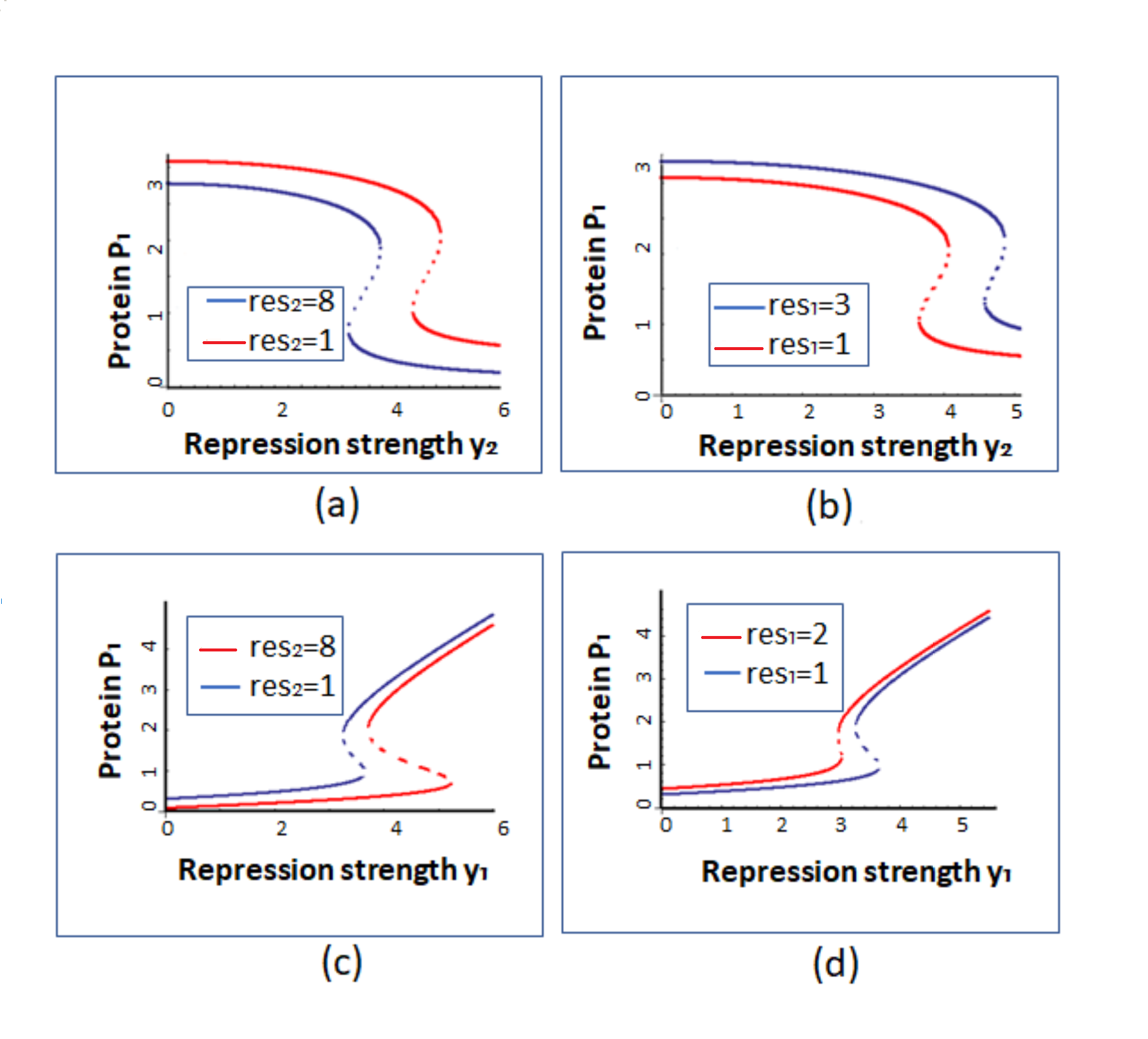}
\end{center}
\end{minipage}
\caption{Linear shift of region of bistability: (a) We depict the bistable response $p_1$ w.r.t. $y_2$. There is a shift of bistable region for different $res_2$ value. $y_2-p_1$ plot keeping $res_1=1, y_1=3.5 ,\delta c_1= \delta c_2=2.5, \delta p_1= \delta p_2= 0.3, \epsilon_1= \epsilon_2=1, g_1= g_2=5$  fixed and $y_2$ is changed from $0$ to $6$. For low resource to $p_2$, $res_2=1$ the red curve describes the nature, $res_2=8$ is for the blue curve. (b) We see in the $p_1$ w.r.t. $y_2$ plot, there is a shift of bistable region for different $res_1$ values. The $y_2-p_1$ plot when $res_2=1, y_1=3.5,\delta c_1= \delta c_2=2.5, \delta p_1= \delta p_2= 0.3, \epsilon_1= \epsilon_2=1, g_1= g_2=5$  is fixed and $y-2$ is varied from $0$ to $5$ is shown. For low resource to $p_1$, $res_1=1$ red curve describes the nature, for comparatively greater resource to $p_1$, $res_1=3$ blue curve is obtained. (c) $y_1-p_1$ plot with $res_1=1$ and the rest fixed as (a),$y_1$ is varied from $0$ to $6$, blue curve is for $res_2=1$ and red curve is for $res_2=8$ (d) $y_1 vs p_1$ plot with $res_2=1$, and rest parameters are same as (a),  $y_1$ is changed from $0$ to $5$, blue curve is for $res_1=1$ and red curve describes the nature when $res_1=2$}
\label{shift}
\end{figure}
\subsection{Resource competition in protein synthesis}
We start by focusing on the effects of asymmetric resource affinities  on $p_1$ which is of fundamental importance for this motif. Fig. \ref{satur}(a)-(b) depicts the dependence of $p_1$ on $res_1$ and $res_2$ respectively for a fixed value of all the other parameters including  $y_1$ and $y_2$. Fig. \ref{satur} (a) shows that $p_1$  increases with growing  $res_1$, which can be explained straightforwardly based on higher resource affinity resulting into higher synthesis.  Fig. \ref{satur}(b)  however is more fascinating; as the total resource is fixed, a greater $res_2$ value restricts the the number of free  ribosomes available to $g_1$. Now if resource available for synthesis of $p_1$ is low, production will automatically decrease, so for the same parameter values, $p_1$ is less when $res_2$ is high.
\begin{figure}
\begin{minipage}[c]{0.99\linewidth}
\begin{center}
\includegraphics[width=13.2cm]{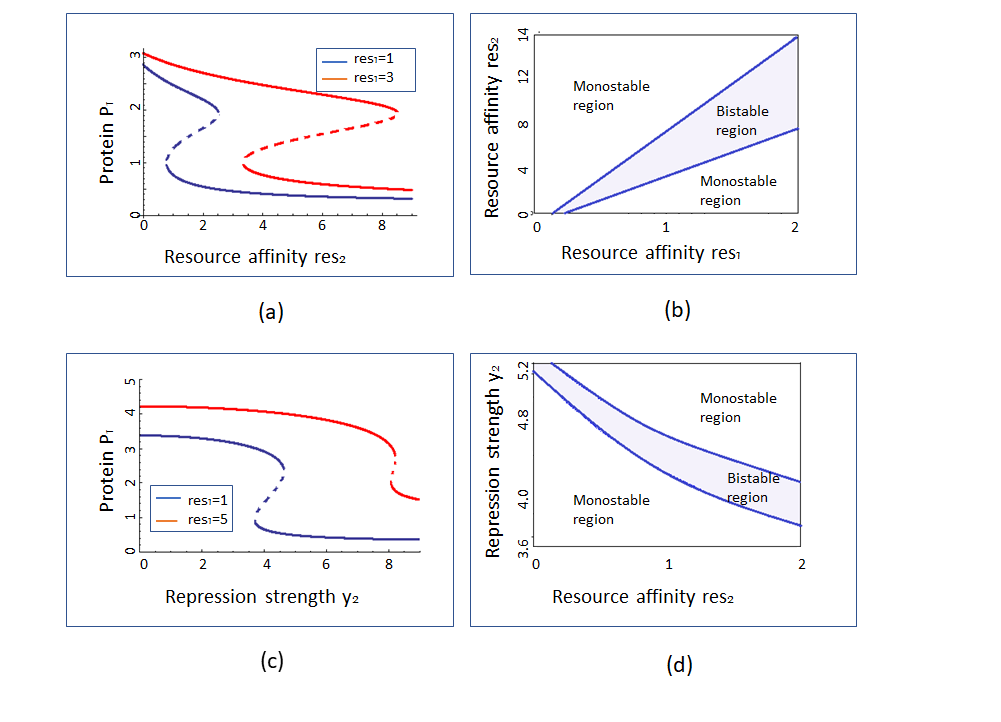}
\end{center}
\end{minipage}
\caption{Effect on functionality of the gene circuit and phase diagrams. (a) Difference between between resource affinities can shift the region of interest, creating monostability in place of bistability for relevant parameter regime. In the $p_1-res_2$ plot, with $y_1=y_2=3.5,T=1,\epsilon_1=\epsilon_2=1,\delta c_1=\delta c_2=2.5, \delta p_1=\delta p_2=0.3,g_1=g_2=5$ fixed, varying $res_2$ from $0$ to $8$, the red curve is for $res_1=3$ and the blue curve is for $res_1=1$. (b) Resource affinity $res_1-res_2$ phase plot, with $y_1=3.5, y_2=3.5,\epsilon_1=2,\epsilon_2=2$ fixed. (c) Extreme disparity between resource affinities destroy the effective region of interest.$y_2-p_1$ plot, with $y_1=4,\epsilon_1=\epsilon_2=1,\delta c_1=\delta c_2=2.5, \delta p_1=\delta p_2=0.3,g_1=g_2=5$ fixed and $y2$ is varied from $0$ to $6$ (d) Phase diagram in $res_2-y_2$ space. $res_1=1,y_1=3.5,\epsilon_1=2,\epsilon_2=2$ is fixed, $res_2$ is changed from $0$ to $2$.}
\label{bist}
\end{figure}
\subsection{Deviation from Proportional Behaviour of Downstream Protein}
We have seen that sharing and competition of resource restricts the linear dependence of $p_2$ on $p_1$ in a region beyond which the $p_2$ level is more controlled by resource competition rather than $p_1$ level and not directly proportional to $p_1$ anymore.  Instead, we observe a restricted region where the proportional nature of response holds. Fig. \ref{satur}(c)  shows the results related to this. Here, it is shown that the graphical behavior can be segmented prominently in two parts: first, a $p_1$ vs $p_2$ linear region which is followed by a saturation effect is which is also very prominent.
The blue curve is when $res_2=0.1$ and $res_1=1$, $p_2$ is nearly constant w.r.t. $p_1$; this is happening because $p_2$ is not being able to gather enough resource for its production, so even if $p_1$ is increasing, $p_2$ cannot increase. The red curve is for intermediate $res_1 (=1)$, where a greater response is received. As $p_2$ can compete for resources now with a comparable value of $res_2$, a proportional nonlinear increase with $p_1$ is observed, which saturates afterwards. The slope gets steeper for greater $res_2$ (green and yellow curve respectively). Thus the availability of resource to $p_2$ controls the span of the region of proportional response.
\subsection{Gradual shift of Region of interest}
The foremost observation after consideration of resource competition is the shift of bistable region in parameter space. In Fig. \ref{shift}(a), we demonstrate that there is a shift of region of interest as we keep all the other parameter fixed and change $res_2$ to observe the effect. A greater $res_2$ value shifts the bistable region to left, causing the bifurcations at lower $y_2$ values. For similar reasons the region of bistability shifts to the right if $res_1$  is increased with other parameters fixed in Fig. \ref{shift}(b). When studied in terms of $y_1$ instead of $y_2$, shift of bistable region is consistently obtained (Fig. \ref{shift}(c)-(d)) and same arguments can be used to explain these results. \\
In Fig. \ref{bist}(b), we elaborate on this observation in $res_2-res_1$ parameter space. Greater $res_1$ causes the bistable region to shift at greater $res_2$ values. This can be qualitatively understood on the basis of correlations between the genes. Greater $res_1$ increases the affinity of $p_1$ gene with resource, producing more $p_1$. Thus, a proportionally higher $p_2$ is expected. For production of high $p_2$, high $res_2$ is required; so the bistability occurs for higher values in $res_2$ parameter regime.

\subsection{Disruption of Switching Dynamics}
Another important observation in this context is disruption of bistablily. If we focus on the region of bistability w.r.t. the $res_2$ parameter, we see for a change in value of $res_1$, bistable region can completely be shifted from the previous location.  In Fig. \ref{bist}(a), with a change of $res_1$ from $1$ to $3$, a slow gradual response can be observed right at the point where previously the bifurcation points existed. Similar results can be observed in Fig. \ref{bist}(c); moreover the region of bistability w.r.t. repression strength becomes extremely small (Fig. \ref{bist}(c)) for higher  $res_1$ values. To observe the correlation between the toggle repression parameters and resource affinity constants, we show the phase diagram in $y_2-res_2$ space in Fig. \ref{bist}(d). We highlight that for low values of $res_2$ the region of bistability shifts very swiftly, and with moderate changes in $res_2$, the previous bistable region turns into monostable region disrupting the switch response completely.

\section{Concluding Remarks}
The interplay between multiple participants in a functional genetic motif imposes substantial challenges in experimental observations and controlled \textit{in-vitro} observations. When the supply pool of available resources heavily exceeds the demand for gene expression, the synthesis process of various motifs in a network can be considered independent. However, as each of these resources is present at a finite number in the cell, unprecedented competitions arise as soon as the network moves on to a resource limited phase. In this paper we propose a model to understand resource trade-off driven coupling between proteins of the same pathway that can affect the circuit functionality. Starting with the model of two genes connected by a positive regulation, we build up a mechanistic model to show the effect of depletion of resources, especially in the context of translational resource, ribosomes.  With extensive parameter sweeps and simulations, we explore how circuit parameters affect the isocline that relates the expression of both proteins. We focus on the impact of copy number, RBS strength and promoter strength on the slope and intercepts of the isocline. Interestingly we find that as expression levels increase, two regimes emerge, one in which the circuit is under-loaded and the isocline behaves as expected, and an overloaded regime in which the isocline is replaced by a nonlinear relation. We investigate how the tipping point between the two regimes can be controlled with circuit parameters, and how these properties depend on $T$, the parameter representative of the number of ribosome molecules. We gather important observations like loss of positive correlation as a result of this competition, and the results are found to be consistent with statistical mechanical equilibrium models. These results are useful to detect the parameters having the largest impact on competition and provide guidelines to control it. \\
Taking one step further we extend the model for a common multigene motif, genetic toggle, to observe how the bistability of the circuit gets affected and eventually disrupted due to competition for ribosome with a downstream protein. Here we must note that the model is computationally light and extremely flexible which can be further extended to include competition for more than two mRNAs. Though created with ribosome depletion problems, this model can be generalised for expression factors associated with all stages of gene expression: competing transcription factors for DNA binding sites to competing codons for tRNAs. This model takes into account of local competitions considering local densities of binder molecules, concentration of gene expression factors and copy numbers of concerned mRNAs in a simpler simulation framework.  \\
In a future work, we will be developing computational models for resource competitions present in all gene expression steps. Our proposed  model can be extremely versatile; for example, this model can be extended for studying ultrasensitive responses in miRNA mediated effects on gene expression, and can be coupled with existing theoretical and experimental results in post-transcriptional regulation \cite{bose2012origins,koscianska2015cooperation,saito2012target}. Consideration of structural information like 3D shapes and binding configurations for multimers \cite{schmidt2014integrated}, exclusive looping architectures related to simultaneous resource binding \cite{rydenfelt2014statistical} etc. can give rise to further interesting questions. These analyses can reveal novel design principles for synthetic circuits and also develop understanding on principles of robust regulation in natural gene regulatory pathways.

\section*{Acknowledgement}
\noindent PC and SG acknowledge the support  by DST-INSPIRE, India, vide sanction Letter No. DST/INSPIRE/04/2017/002765  dated- 13.03.2019.
\section*{References}
\bibliographystyle{iop}
\bibliography{ws-sample.bib}

\end{document}